\documentclass[twocolumn,showpacs,amsmath,amssymb,dvipdfm]{revtex4}
\usepackage{graphicx,color,multirow,hyperref}

\newcommand{\fcb}{f_{\rm cb}}
\newcommand{\fnu}{f_{\nu}}

\newcommand{\bfp}{\mbox{\boldmath$p$}}

\newcommand{\bmf}[1]{\mbox{\boldmath$#1$}}
\newcommand{\be}{\begin{equation}}
\newcommand{\ee}{\end{equation}}
\newcommand{\ba}{\begin{eqnarray}}
\newcommand{\ea}{\end{eqnarray}}

\newcommand{\ltsim}{\protect\raisebox{-0.5ex}
  {$\:\stackrel{\textstyle <}{\sim}\:$}}

\newcommand{\simgt}{\lower.5ex\hbox{$\; \buildrel > \over \sim \;$}}
\newcommand{\simlt}{\lower.5ex\hbox{$\; \buildrel < \over \sim \;$}}

\begin{document}


\title{%
Neutrino mass constraint with the Sloan Digital Sky Survey power spectrum of luminous red galaxies and perturbation theory
}%
\author{%
 Shun Saito$^{1,2}$, Masahiro Takada$^{3}$ and Atsushi Taruya$^{3,4}$
}%
\affiliation{%
 $^1$Department of Physics, School of Science, 
 The University of Tokyo, Tokyo 113-0033, Japan
}%
\affiliation{%
 $^2$Department of Astronomy, 
 University of California at Berkeley, 601 Campbell Hall, Berkeley, California 94720, USA
}%
\affiliation{%
 $^3$Institute for the Physics and Mathematics of the Universe (IPMU), 
 The University of Tokyo, Chiba 277-8582, Japan
}%
\affiliation{%
 $^4$Research Center for the Early Universe, 
 School of Science, The University of Tokyo, Tokyo 113-0033, Japan
}%

\date{\today}

\begin{abstract}
We compare the model power spectrum, 
computed based on 
perturbation theory
with the power spectrum
of luminous red galaxies (LRG) measured from the Sloan Digital Sky
Survey Data Release 7 catalog, assuming 
a flat, cold dark matter-dominated cosmology. 
The model includes the effects of massive neutrinos,
nonlinear matter clustering and nonlinear, scale-dependent galaxy bias
in a self-consistent manner. 
We first test the accuracy of perturbation theory model by comparing the model
predictions with the halo power spectrum in real- and redshift-space,
measured from 70 simulation realizations for a cold dark matter model without massive
neutrinos. We show that the perturbation theory model with bias parameters being properly
adjusted can fairly well reproduce the simulation results. As a result
the best-fit parameters obtained from the hypothetical parameter fitting
recover, within statistical uncertainties, the input cosmological
parameters in simulations, including an upper bound on neutrino mass, if
the power spectrum information up to $k\simeq 0.15~h$Mpc$^{-1}$ is
used. However, for the redshift-space power spectrum, the best-fit
cosmological parameters show a sizable bias from the input values if
using the information up to $k\simeq 0.2~h$Mpc$^{-1}$, probably due to
nonlinear redshift distortion effect.  Given these tests, we decided,
as a conservative choice, to use the LRG power spectrum up to
$k=0.1~h$Mpc$^{-1}$ in order to minimize possible unknown nonlinearity
effects.
In combination with the recent results from Wilkinson Microwave
Background Anisotropy Probe (WMAP), we derive a robust upper-bound on
the sum of neutrino masses, given as $\sum m_{\nu}\le 0.81~{\rm eV}$
(95\% C.L.), marginalized over other parameters including nonlinear
bias parameters and dark energy equation of state parameter. The upper
bound is only slightly improved to $\sum m_{\nu}\le 0.80~{\rm eV}$ if
including the LRG spectrum up to $k=0.2~h$Mpc$^{-1}$, due to severe
parameter degeneracies, though the constraint may be biased as 
discussed above.  The neutrino mass limit is improved by a factor of 
1.85 compared to the limit from the WMAP5 alone, $\sum m_\nu\le 
1.5~{\rm eV}$.
\end{abstract}


\maketitle

\section{Introduction} 
Combining the cosmic microwave background (CMB) 
with large-scale structure probes provides a powerful means of constraining the sum of neutrino
masses \cite{Huetal:1998,Takadaetal:2006}. 
Massive neutrinos imprint a characteristic suppression in the clustering
of galaxies at scales 
below the free-streaming scale of neutrinos in a cold
dark matter  (CDM) dominated structure formation scenario.  
In particular, for neutrino masses of ${\sim}0.1$eV inferred from
terrestrial experiments, a wide-field galaxy redshift survey can directly
probe the scales comparable with the neutrino free-streaming scale
${\sim}100$Mpc, which is incidentally close to the baryonic acoustic
oscillation scales. The existing galaxy surveys
have provided a stringent
upper limit on the neutrino mass
(\cite{Elgaroy:2002lr,Tegmark:2006fk,Thomas:2010lr}; see Reid et al. 2010  
\cite{Reid:2010qy} 
for the most recent study, hereafter R10). 
However, all the previous studies employed a rather empirical approach to model the
nonlinear effects in galaxy clustering such as the $Q_{\rm nl}$-model 
\cite{Tegmark:2006fk}
or the method to use polynomial functions of
wavenumbers with additional nuisance parameters \cite{Reid:2010qy}.
We also note that these modelings have been tested using mock catalogs
without neutrino effects being taken into account.

In order to derive a robust, reliable constraint on 
neutrino masses from the observed galaxy distribution, an
accurate modeling of galaxy clustering is clearly needed
properly taking into account
the effects of nonlinear clustering, redshift distortion and
nonlinear, scale-dependent galaxy bias.
Simulation based approach 
may be the most powerful method, however, such a study for a mixed
dark matter model (CDM plus massive neutrinos) is still in 
developing stages, 
especially for small neutrino mass scales of interest
$\ltsim 1~{\rm eV}$ \cite{Brandbyge:2009,Vieletal:2010,Feldman:2010}. An
analytical approach is complementary, and allows us to study the effect
of massive neutrinos as a function of different cosmological models.
Recently we have developed the new analytical method
to compute 
the nonlinear galaxy power spectrum based on the perturbation theory (PT) approach 
\cite{Saito:2008lr,Saito:2009fk}, where the effects of
nonlinear clustering and nonlinear galaxy bias are included in a
self-consistent manner within the PT framework. 
The PT-based model (see also \cite{Wong:2008qy,Shoji:2009fk,Lesgourgues:2009uq}) 
is a natural extension of the well-established linear theory, and the
validity has been extensively studied by comparing with {\it N}-body
simulations in a CDM model
(e.g. \cite{JeongKomatsu:2009,Nishimichietal:2009,Taruya:2009uq,Taruya:2010fj}).

In this paper, we present the first application of the PT model to 
the power spectrum of luminous red galaxies (LRGs)
measured from the Data Release 7 catalog of the Sloan Digital Sky
Survey (SDSS) in R10. We then derive a robust constraint on neutrino
masses, combined with the WMAP 5-year (WMAP5) data
\cite{Komatsu:2009qy}, including 
marginalization over the uncertainties of galaxy bias parameters, residual shot noise
contribution, and dark energy parameters. 
We mention the recent study \cite{Swanson:2010lr}, where the PT-based model
 is compared to the SDSS main galaxies (not LRGs). 

\section{Modeling of nonlinear galaxy power spectrum} 

\subsection{PT model} 

In our previous papers \cite{Saito:2008lr,Saito:2009fk} 
we developed a PT-based method 
for computing the nonlinear galaxy power spectrum in a
mixed dark matter model: 
\begin{equation}
P_g(k; z) = b_{1}^{2}\left[ 
 P^{\rm NL}_{\rm m}(k; z)+b_{2}P_{\rm b2}(k;z)+b_{2}^{2}P_{\rm b22}(k;z)
 \right] + N.
\label{eq:nonlinearPkhalo}
\end{equation}
Here $b_1$ and $b_2$ are the linear and nonlinear bias parameters and
$N$ denotes the residual shot noise parameter, which are derived 
by renormalizing the galaxy bias parameters based on PT prescription \cite{McDonald:2006kx}. 
The expressions for power spectra $P^{\rm NL}_{\rm m}$, 
$P_{\rm b2}$ and $P_{\rm b22}$ are given in \cite{Saito:2009fk}. Note that 
$P_{\rm b2}>0$ and $P_{\rm b22}<0$ over a range of relevant scales. 
The nonlinear matter power spectrum $P^{\rm NL}_{\rm m}$ is given as 
\ba 
&&P^{\rm NL}_{\rm m}(k) = \fcb^{2} [{P^{\rm L}_{\rm
 cb}}(k)+{P^{(22)}_{\rm cb}}(k)+{P^{(13)}_{\rm cb}}(k)]\nonumber\\
 &&~~~~~~~~~~~~ +2\fcb\fnu P^{\rm L}_{\rm cb\nu}(k) + \fnu^{2} P^{\rm
 L}_{\nu}(k), \label{eq:nonlinearPkmatter}
\ea
where the subscripts ``cb'' and ``$\nu$'' denote ``CDM plus baryon'' and
``massive neutrinos,'' respectively, and $P^{(13)}_{\rm cb}$ and
$P^{(22)}_{\rm cb}$ 
describe the perturbative corrections to the power spectrum 
at next-to-leading order. 
The coefficient $f_i$ is the mass
fraction of each species relative to the present-day energy density of
total matter, $\Omega_{\rm m0}$: 
$f_{\nu}\equiv \Omega_{\nu0}/\Omega_{\rm m0} =\sum
m_{\nu}/(\Omega_{\rm m0} h^2 
\times 94.1{\rm eV})$ and $f_{\rm cb}=1-f_\nu$.
The nonlinear galaxy power 
spectrum at a given redshift $z$ [Eq.~(\ref{eq:nonlinearPkhalo})] can be
computed once the linear-order power spectra of CDM, baryon and neutrino
perturbations at the same redshift $z$ are given for an assumed cosmological
model and the bias parameters $b_1$, $b_2$ and $N$ are specified, 
as extensively studied in \cite{Saito:2009fk}.

\subsection{Testing PT model with simulations} 

To assess the validity of the PT model [Eq.(\ref{eq:nonlinearPkhalo})]
in estimating model parameters, we implement a hypothetical experiment:
By fitting the PT model to the halo power spectrum measured from {\it N}-body
simulations, we address whether the cosmological parameters assumed in
{\it N}-body simulations can be properly recovered.  We used 70 simulation
realizations each of which is carried out with 512$^3$ {\it N}-body particles
and volume of 1~$h^{-3}$Gpc$^3$, comparable with the volume covered by
the SDSS survey (the {\it N}-body simulations are kindly provided by Issha
Kayo, and also see \cite{Tangetal}).  We created the halo
catalogs from {\it N}-body simulation outputs at $z=0$, 
based on the friend-of-friend method with the linking length of
$b=0.2$ (20\% of the mean separation of {\it N}-body particles). 
The catalog in each realization contains halos with masses greater than 
$M_{\rm min}\simeq 10^{13}h^{-1}M_\odot$, 
where the 
mass threshold is determined 
such that the resulting 
number density of halos becomes $\bar{n}_{\rm halo}\simeq 3\times
10^{-4}h^3$Mpc$^{-3}$, 
comparable with that of the SDSS LRGs.

\begin{figure}[t]
\begin{center}
\includegraphics[width=0.45\textwidth]{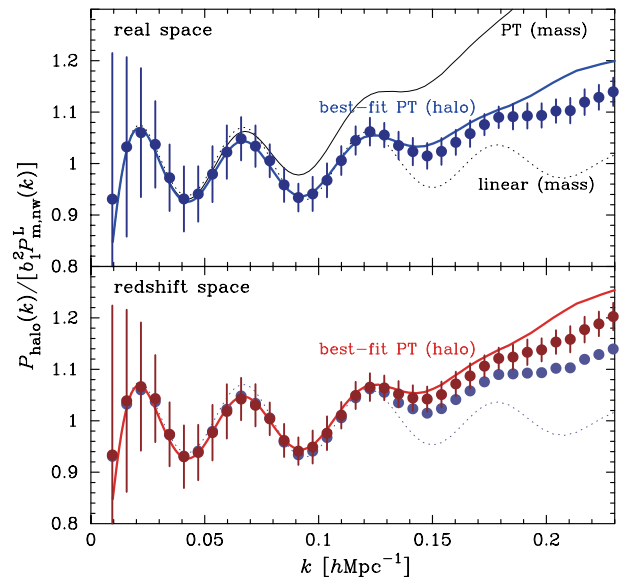}
\end{center}
\vspace*{-2em}
\caption{ {\em Upper panel}: The filled circles at each $k$ bin show the
mean halo power spectrum measured from 70 simulation realizations at
$z=0$ (see text for details), while the error bar shows the statistical
measurement uncertainty at the $k$ bin for a simulation volume of
1~$h^{-3}$Gpc$^3$, roughly comparable with the SDSS survey volume. For
illustrative purpose the halo spectrum is divided by the no-wiggle,
linear power spectrum, multiplied by the linear halo bias squared,
$b_1^2 P^L_{{\rm m,nw}}(k)$ ($b_{1}=1.66$). 
For comparison, the thin-dotted and -solid curves show the linear-theory
and PT predictions for ``mass'' power spectrum, respectively, for the
cosmological model assumed in the simulations. 
The bold solid curve shows the best-fit PT model for halo power
spectrum, computed from Eq.~(\ref{eq:nonlinearPkmatter}), where the
best-fit model parameters including bias parameters are obtained by
fitting the model predictions to the simulation spectrum up to
$k=0.1~h$Mpc$^{-1}$ (see Fig.~\ref{fig:HaloTest_1D}).  
{\em Lower panel}: Similar to the upper panel,
but for redshift-space power spectrum ($b_{1}=1.81$). 
The redshift-space power spectrum is modified by redshift
distortion effect due to peculiar velocities of halos. For comparison,
the circle points without error bars show the simulation halo spectrum
in real space (the same as in the upper panel).
}  
\label{fig:RvsS}
\end{figure}

Figure~\ref{fig:RvsS} shows the halo power spectra in real- and redshift
space, measured from the 70 simulation realizations above. The
redshift-space power spectrum shown here is the monopole spectra,
i.e. obtained by azimuthally averaging the 2D power spectrum in redshift
space over circular annulus of a given radius of $k$, where the
line-of-sight direction is taken as the $z$-direction of each simulation
box.  
The filled circle
at each $k$ bin shows the mean band power computed from the 70
realizations, while the error bar is the scatter corresponding to the
statistical measurement uncertainty for a survey volume of
$1~[h^{-1}{\rm Gpc}]^3$. The upper and lower panels show the real- and
redshift-space power spectra, respectively, where the redshift-space
power spectrum is affected by redshift distortion effect due to peculiar
velocities of halos. The redshift distortion effect considered here
arises from the bulk motion of halos, because the peculiar velocity of
each halo in simulation is defined by the mean of velocities of member
{\it N}-body particles in the halo and therefore the internal virial motions
are averaged out.  The linear theory predicts that the redshift
distortion effect by the bulk motion causes only an overall shift in the
power spectrum amplitude \cite{Kaiser:87}, and does not change the shape
of power spectrum. 
However, as can be found from the lower panel, the
redshift-space halo spectrum, which is normalized so as to match the
real-space power spectrum amplitude at small $k$, shows a
scale-dependent enhancement in the amplitudes compared to 
the real-space spectrum.  This implies nonlinear redshift distortion due
to the halo velocity field. (Also see \cite{Hamanaetal:03} for discussion
on velocity bias of halos.) 
Recently, the authors of \cite{Taruya:2010fj} developed a model to compute the
nonlinear, redshift-space power spectrum taking into account
nonlinearity effects such as mass clustering and redshift distortion,
based on perturbation theory.  Interestingly, this work showed that such
a scale-dependent enhancement in the redshift-space power spectrum
amplitude may be caused by the nonlinear peculiar velocity, which
qualitatively explains the results shown in Fig. \ref{fig:RvsS}.
However, a further exploration of this effect is beyond the scope of
this paper, and will be studied elsewhere. 

In Fig.~\ref{fig:RvsS} we compare the halo spectra, measured from the
simulations, with the model predictions computed based on linear theory
or the PT model. For the PT predictions, we show the mass power spectrum
(thin solid curve) assuming the cosmological model assumed in the
simulations, and also show the best-fit halo power spectrum (bold solid
curve), computed from Eq.~(\ref{eq:nonlinearPkmatter}), where the
best-fit model parameters including bias parameters are obtained by
fitting the PT model to the simulation spectrum up to
$k=0.1~h$Mpc$^{-1}$ (also see below for details). The figure clearly
demonstrates that the simulation halo spectra cannot be explained by
either the linear theory or the PT model for mass power spectrum over a
range of wavenumbers of interest. Although the linear theory may appear
to give a good fit to the simulation result up to $k\simeq
0.12~h$Mpc$^{-1}$,  a closer look reveals that the linear theory
over-estimates the band powers at  scales around
$k\simeq 0.07~h$Mpc$^{-1}$, where the PT model gives a better fit.  On
the other hand, interestingly, the PT-based halo spectrum with bias
parameters being properly adjusted fairly well reproduces the simulation
results. The PT model can give a reasonably good fit up to $k\simeq
0.15~h$Mpc$^{-1}$, but begins to increasingly deviate from the
simulation results at the larger $k$ due to stronger nonlinearity
effects.

Next let us move on to details of the hypothetical parameter fitting;
can the PT model for halo power spectrum recover the cosmological
parameters assumed in the simulations?  In the parameter fitting, we
employ 6 parameters: $\Omega_{\rm b0}/\Omega_{\rm m0}(=0.172)$,
$\Omega_{\rm m0}h(=0.174)$, $\sum m_{\nu}(=0~{\rm eV})$, and the
parameters of galaxy bias and shot noise, $b_1$, $b_2$ and $N$. (The
values in parentheseis are the input values assumed in the {\it N}-body
simulations.) Other parameters are fixed to the input values of {\it N}-body
simulations.  Note that adding the neutrino mass as a free parameter
causes an asymmetric effect on the model spectra due to the sharp limit
$\sum m_\nu\ge 0$: it causes only a scale-dependent suppression, not an
increase, in the power spectrum amplitudes.  We imposed the Gaussian
prior $\sigma(\Omega_{\rm b0})/\Omega_{\rm b0}=0.05$, and
assumed the residual shot noise $N$ 
to be smaller than the 
power spectrum amplitudes over the range of wavenumbers used. Finally,
we used the power
spectrum information up to the maximum wavenumber $k_{\rm
max}=0.1~h{\rm Mpc}^{-1}$, motivated by the fact that the PT model
stays fairly accurate down to the wavenumber $k\lesssim0.1~h$Mpc$^{-1}$ 
for {\em mass} power spectrum at $z=0$ 
in a CDM model as carefully studied in
 \cite{Taruya:2009uq,Nishimichietal:2009}.

\begin{figure}[t]
\begin{center}
\includegraphics[width=0.48\textwidth]{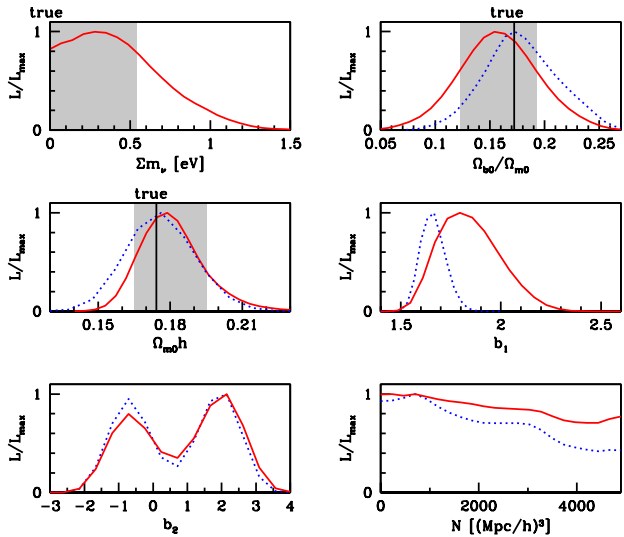}
\end{center}
\vspace*{-2em}
\caption{ Testing the perturbation theory (PT) based model
with the halo power spectrum measured from {\it N}-body simulations 
(70 realizations used). 
The solid curve in each panel is the posterior distribution of parameter, 
estimated by comparing the PT model with the halo power spectrum up to the
maximum wavenumber $k_{\rm max}=0.1~h{\rm Mpc}^{-1}$.  The input values
of $\Omega_{\rm b0}/\Omega_{\rm m0}$ and $\Omega_{\rm m0}$, denoted by
the vertical lines, are properly recovered within the statistical errors
for the volume $1~h^{-3 }{\rm Gpc}^3$, which is comparable with the SDSS
volume. For neutrino masses, which are not included in the {\it N}-body
simulations, an upper limit is derived.  Nonzero values of bias
parameters ($b_1$ and $b_2$) and shot noise parameter ($N$) are
obtained, implying that the parameters are needed to describe the halo
power spectrum. The dotted curves represent the posterior distribution 
obtained by fixing the neutrino mass to zero, showing that    
the input values of the parameters $\Omega_{\rm b0}/\Omega_{\rm m0}$ 
and $\Omega_{\rm m0}$ are correctly reproduced 
together with a tighter constraint on the linear bias parameter. 
} \label{fig:HaloTest_1D}
\end{figure}

Figure~\ref{fig:HaloTest_1D} shows the resulting posterior distribution
of each model parameter.
It is found that the input values of
$\Omega_{\rm b0}/\Omega_{\rm m0}$,
$\Omega_{\rm m0}h$  and $\sum m_\nu$
are well recovered within the 68\% C.L. statistical
uncertainties as denoted by the shaded regions. 
A closer look reveals that 
the best-fit values 
slightly deviate  
from the input values.
The origin of the offsets can be 
explained by the dotted curves, which show
the posterior distribution obtained by 
fixing the neutrino mass to zero (i.e., the input value for the CDM 
simulations). 
Adding neutrino masses as a free parameter causes
such a bias in the best-fit values 
of $\Omega_{\rm b0}$, $\Omega_{\rm m0}h$ and $b_1$.
This bias direction in 
$\Omega_{\rm b0}$, $\Omega_{\rm m0}h$ and $b_1$ 
is found to all increase 
the power spectrum amplitudes
so as to
compensate a scale-dependent suppression caused 
by nonzero neutrino masses.
Accordingly an upper limit on $\sum m_\nu$ is obtained due to 
the sharp cutoff $\sum m_\nu\ge 0$. It should also 
be noted that adding neutrino masses increases
the marginalized error of $b_1$, 
implying a strong degeneracy between $b_1$
and $\sum m_\nu$. 

A nonzero value of $b_2$ 
is favored for the PT model to match the halo power spectrum or
equivalently a simple linear bias is disfavored even 
for $k_{\rm max}=0.1~h$Mpc$^{-1}$, as also inferred from Fig.~\ref{fig:RvsS}. 
The bimodal distribution of $b_2$ is also apparent. 
For the favored values of $b_2$,
the term proportional to $b_2^2$ in Eq.~(\ref{eq:nonlinearPkhalo}) is
dominant over the term proportional to $b_2$, and therefore both
positive and negative values of $b_2$ become acceptable. 

How is the parameter estimation changed if using the PT model up to
higher $k_{\rm max}$ than 0.1~$h$Mpc$^{-1}$, where the PT model ceases
to be accurate, at least for the mass power spectrum
\cite{Taruya:2009uq,Nishimichietal:2009}? 
Some of the previous works sometimes attempted to use the power spectrum
information up to such higher $k$-range, 
motivated by the fact that the power
spectrum of higher-$k$ modes contains a more constraining power of
parameters. 
However, because of complex nonlinearity effects, the best-fit
parameters derived from such high-$k$ information may be biased from the
underlying true values.  On the other hand, the PT model predicts a
complex scale-dependent, nonlinear bias function as a function of
cosmological model and bias parameters (see \cite{Saito:2009fk}). 
As implied in Fig.~\ref{fig:RvsS}, the PT model can give a good fit
to the simulation halo spectrum at scales greater than
$k=0.1~h$Mpc$^{-1}$, 
by  adjusting the bias parameters. Therefore it is interesting 
to study whether or not 
fitting the PT model to the simulation halo
spectrum up to the higher $k$-values causes a bias in the best-fit parameters. 

\begin{figure}[t]
\begin{center}
\includegraphics[width=0.48\textwidth]{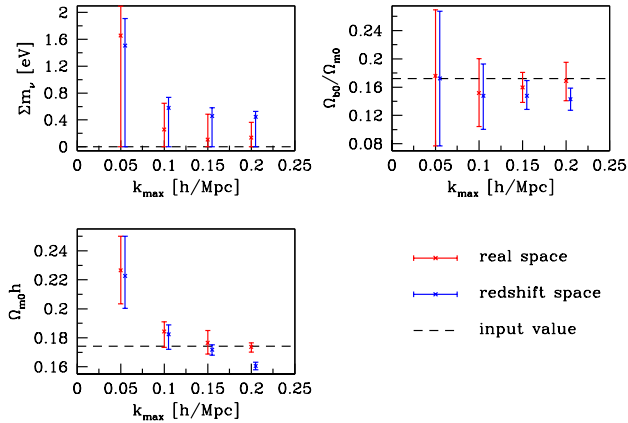}
\end{center}
\vspace*{-2em}
\caption{The best-fit parameters and the marginalized errors obtained by
 fitting the PT model with the halo spectrum up to a given maximum wavenumber 
 $k_{\rm max}$, denoted in the horizontal axis. For each $k_{\rm max}$
 the left-side point with error bar shows the results for the real-space halo
 spectrum, while the right-side point shows the results for redshift-space
 halo spectrum. 
 The horizontal dashed line denotes the input value of each parameter. 
} 
\label{fig:Fig1D_kmax}
\end{figure}

Figure~\ref{fig:Fig1D_kmax} shows the 1$\sigma$
marginalized errors on $\Sigma m_{\nu}$, $\Omega_{\rm b0}/\Omega_{\rm
m0}$ and $\Omega_{\rm m0}h$ as a function of the maximum wavenumber
$k_{\rm max}$ employed in the parameter fitting.  First, let us focus on
the results for the real-space power spectrum.  Even though the PT model
breaks down at $k\simgt 0.1h$Mpc$^{-1}$ and over-estimates the ``mass''
power spectrum amplitudes at such high-$k$ range
(see the upper panel of Fig.~\ref{fig:RvsS}), 
the best-fit parameters are found to recover the input values within the
$1$-$\sigma$ statistical uncertainties. It is also clear that the
statistical errors of the parameter and the upper bound on $\Sigma
m_\nu$ are improved at $k_{\rm max}\ge 0.1~h^{-1}$Mpc$^{-1}$ compared to
our fiducial choice of $k_{\rm max}=0.1~h$Mpc$^{-1}$, due to a gain in
the constraining power contained in the high $k$-range. This may reflect
that the PT model has more degrees of freedom to describe the nonlinear
halo power spectrum by adjusting the bias parameters, which may allow one to
overcome the limitation of PT model for mass power spectrum. 

However, this is not the case for the redshift-space
halo power spectrum. Again note that, if the halo power spectrum measured
from simulations is affected only by the Kaiser effect, the redshift
distortion effect causes only an overall shift in the power spectrum
amplitude, independent of $k$, which can be absorbed by changing the
linear bias parameter in the PT modeling. The figure shows that the
input parameters are recovered up to $k_{\rm max}\simeq
0.15~h$Mpc$^{-1}$, but the best-fit parameters at $k_{\rm
max}=0.2~h$Mpc$^{-1}$ show a sizable deviation, more than the
statistical uncertainties, compared to the input values for $\Omega_{\rm
m0}h^2$ and $\Omega_{\rm b0}/\Omega_{\rm m0}$.  This deviation implies
that the residual nonlinear redshift distortion cannot be described by
the PT model, even if changing the model parameters.

Note that, on the contrary, 
at low $k_{\rm max}\sim$0.05$~h$Mpc$^{-1}$, there are less statistical
powers, giving larger uncertainties in model parameters. 
In addition, severe parameter degeneracies give only a very weak upper
bound on neutrino mass, and cause a bias in $\Omega_{\rm m0}h$ as
discussed above. 

Given the results in Figs.~\ref{fig:RvsS} and
\ref{fig:HaloTest_1D}, we will use, as a conservative choice, the SDSS
LRG power spectrum up to $k=0.1~h$Mpc$^{-1}$ to compare with the PT
model. For the LRG power spectrum measured in R10, the nonlinear
redshift distortion, known as the Fingers-of-God effect,
 is suppressed  by clipping possible satellite LRGs. However, as we
have shown, there may be a residual contamination from the nonlinear
redshift distortion effect. Therefore, in order to derive a robust
constraint on neutrino mass, we will adopt $k_{\rm max}=0.1~h$Mpc$^{-1}$
for the following results, although we will also discuss how the
neutrino mass constraint is changed by including the information beyond
$k=0.1~h$Mpc$^{-1}$, to be more comprehensive.

\section{Results} 
We now apply the PT model to the power spectrum 
of SDSS LRG samples 
in order to constrain neutrino mass. 
We use the halo power spectrum measured by R10, 
where 104,337 halos were first reconstructed from the observed 110576 LRGs' 
distribution, and the angle-averaged redshift-space power spectrum was 
estimated based on the method in \cite{Percival:2004np}.
The halo power spectrum is less 
affected by the Fingers-of-God effect, 
because the contribution from satellite galaxies was eliminated in the measurement. 
Thus the halo power spectrum in R10 is appropriate to compare with the PT model.  

In estimating parameters, we combine the LRG power spectrum 
with the WMAP5 data. 
Note that our results would remain almost unchanged even with the latest 
WMAP7 result \cite{Komatsu:2010yq}. 
We assume that the likelihood function of the LRG power spectrum is given as
\be
 -2\ln {\cal L}_{\rm SDSS} \propto \sum_{k_{i,j}<k_{\rm max}} \Delta_{i}
[{\bmf{C}^{-1}}]_{ij}\Delta_{j}, 
\ee
where $\Delta_i$ is the difference 
between the measured and model 
power spectra at the $i$-th wavenumber bin $k_i$, 
$\Delta_{i}\equiv \hat{P}_{\rm halo}(k_{i})-P^{\rm NL}_{\rm halo}(\alpha k_{i} | {\bfp})$, 
with $\bmf{p}$ being a set of model parameters (see below). 
Note that the effect of survey window function is properly taken into 
account in computing the model power spectrum following R10. 
The matrix $\bmf{C}$ is the covariance matrix for which we use
the matrix provided in R10, and $\bmf{C}^{-1}$ is its inverse
matrix. Note that we employ $k_{\rm max}=0.1~h{\rm Mpc}^{-1}$ and assume
the single redshift slice $z=0.35$ for simplicity.
The stretch factor ``$\alpha$'' in the argument of the model power
spectrum describes the cosmological distortion 
\cite{Eisensteinetal:2005,Percival:2010lr}. 
This factor is given 
as $\alpha=D_{V}^{\rm ref}(z)/D_{V}(z; \bfp)$, where 
$D_{V}(z)\equiv [(1+z)^{2}D_{A}(z)^{2}z/H(z)]^{1/3}$
 and $D_V^{\rm ref}$ is the distance for the
reference cosmological model used in the LRG spectrum measurement.  The
likelihood for the joint analysis of WMAP5 plus SDSS is simply given
as $\ln {\cal L}=\ln {\cal L_{\rm SDSS}}+\ln{\cal L}_{\rm WMAP}$. 

We include a fairly broad range of parameters that can cover
variants of 
CDM cosmology
such as models including massive neutrinos and 
dark energy equation of state parameter. 
We vary 12 model parameters in total:
\begin{equation}
\bmf{p}=(f_{\nu},\Omega_{\rm
b0}h^{2},\Omega_{\rm
DM0}h^{2},\theta_{*},w,\tau,\Delta^{2}_{\cal R},
n_{\rm s},A_{\rm SZ}, b_1,b_2,N),
\label{eq:paras}
\end{equation}
where $\Omega_{\rm DM0}h^{2}$ is the sum of CDM and massive neutrinos: 
$\theta_\ast$ is the parameter to characterize the angular scale of CMB
acoustic oscillations: $\tau$ is the optical depth to the last scattering
surface: $n_s$ and $\Delta_{\cal R}^2$ are the parameters to specify the
primordial power spectrum following the convention in
\cite{Komatsu:2009qy}: $A_{\rm SZ}$ is the parameter to control a
contamination of the Sunyaev-Zel'dovich effect to the CMB 
spectrum: $w$ is the dark energy equation of state parameter. Note that
the parameters $\tau$ and $A_{\rm SZ}$ 
affect only the CMB information. 
We used the COSMOMC code \cite{Lewis:2002lr} to explore parameter
estimations in the multidimensional parameter space.

\begin{figure}[t]
\begin{center}
\includegraphics[width=0.48\textwidth]{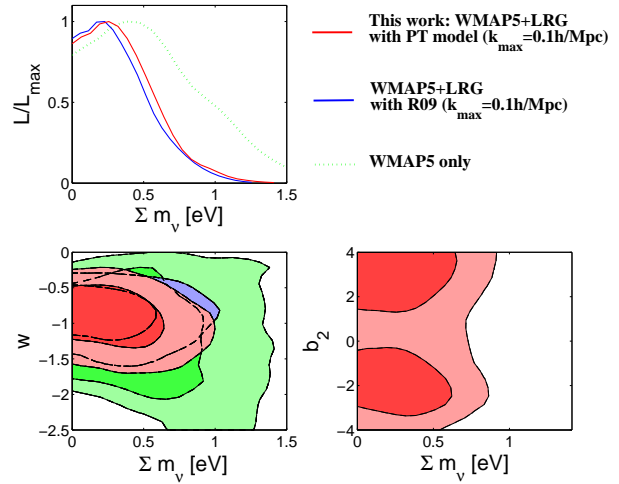}
\end{center}
\vspace*{-2em}
\caption{
 The parameter constraints obtained by comparing the PT model with 
 the SDSS LRG power spectrum up to $k_{\rm max}=0.1~h{\rm Mpc}^{-1}$, 
 in combination with the WMAP5
 constraint, where we include 12 parameters given by Eq.~(\ref{eq:paras}). 
 The upper
 panel shows the posterior distribution of neutrino masses, yielding the upper 
 limit $\sum m_{\nu}\le 0.81~{\rm eV}$ (95\% C.L.), a factor 
 1.85 improvement over the limit $\sum m_\nu\le 1.5~{\rm eV}$ from the WMAP5
 alone.  
 The lower two panels show how the neutrino mass is degenerate
 with the dark energy equation of state parameter $w$ and the nonlinear
 bias parameter $b_2$, respectively, 
 with 68\% C.L. (dark shaded) and 95\% C.L. (light shaded) regions. 
 A nonzero $b_2$ or equivalently a
 scale-dependent bias is favored at 68\% C.L.
 Our results are compared with the results derived using the halo-model
 based  method in
 R10 for the same maximum wavenumber cutoff $k_{\rm max}=0.1~h$Mpc$^{-1}$. 
 }
\label{fig:result}
\end{figure}

\begin{figure}[t]
\begin{center}
\includegraphics[width=0.45\textwidth]{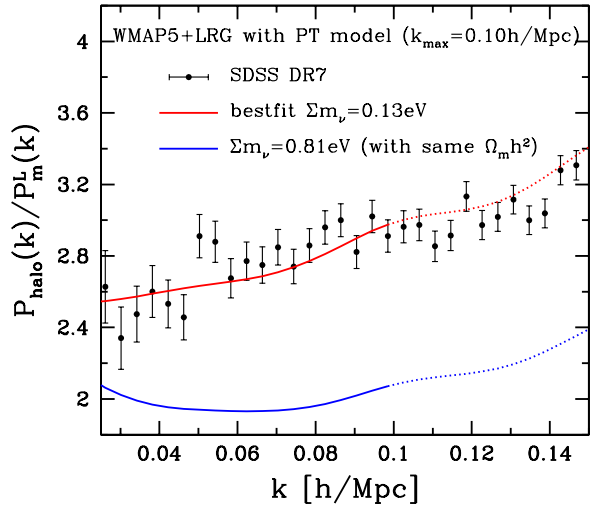}
\end{center}
\vspace*{-2em}
\caption{Comparing the best-fit PT model with the SDSS LRG
 spectrum, where the best-fit model is obtained from the fitting to
 $k_{\rm max}=0.1~h{\rm Mpc}^{-1}$. For illustrative clarity the power
 spectra are divided by the linear matter power spectrum for the
 best-fit cosmological model. 
 For comparison, 
 we also show the PT model, where the neutrino mass is changed to 
 $\sum m_{\nu}=0.81\,$eV, corresponding
 to the 95\% C.L. upper bound in Fig.~\ref{fig:result}, but other parameters
 are kept fixed to the best-fit values. 
 }
\label{fig:PkvsBF}
\end{figure}


The upper panel of Fig.~\ref{fig:result} shows the marginalized error on
neutrino masses. We obtain the upper limit 
$\sum m_\nu\le 0.81~{\rm eV}$
(95\% C.L.) for the SDSS Data Release 7 plus WMAP5. This is 
a factor of 1.85
improvement compared to the limit derived from the WMAP5 alone, 
$\sum m_\nu\le 1.5~{\rm eV}$. 
Our neutrino mass limit can be compared with the result derived using the
method in R10,
where the empirical treatment based on halo model prescription 
was used to account for nonlinear, scale-dependent galaxy bias. 
Note that R10 employed $k_{\rm max}=0.2\,h$Mpc$^{-1}$ 
and then derived an upper bound on the neutrino mass
given as $\sum m_\nu\le 0.63~{\rm eV}$. 

Fig.~\ref{fig:PkvsBF} shows that the best-fit PT model matches the
measured LRG power spectrum well over a range of the working wavenumbers,
$k\le 0.1~h$Mpc$^{-1}$. If the neutrino mass is changed to 
the 95\% C.L. upper bound, 
$\sum m_\nu=0.81~{\rm eV}$, but other parameters are kept fixed to the
best-fit values, the model spectrum significantly underestimates the
measured spectrum amplitudes at the small scales. Also
note that the best-fit model rather continues to match the measured
spectrum beyond $k_{\rm max}=0.1~h{\rm Mpc}^{-1}$. 
In fact, even if including the information up to $k_{\rm max}=0.2~h{\rm Mpc}^{-1}$, 
the neutrino mass limit is only slightly changed to $\sum m_\nu\le 0.8~$eV,
reflecting less cosmological information at the higher wavenumbers due 
to severe degeneracies of cosmological parameters with nonlinear bias 
parameter and/or shot noise parameter. 

The lower two panels of Fig.~\ref{fig:result} show how the neutrino
mass constraint is degenerate with $w$ and the nonlinear bias parameter $b_2$.
The marginalized constraint on $w$ is 
$-1.08<w<-0.79$ (68\% C.L.). 
While a change of $w$ leads to 
a suppression in the power spectrum
amplitudes, Fig.~\ref{fig:result} shows that degeneracy between $w$ and 
the neutrino mass is 
rather weaker than expected. 
This implies that the constraint on $w$ comes
mainly from 
the baryon acoustic oscillation information as studied in \cite{Percival:2010lr}. 
Figure~\ref{fig:result} also shows that
a simple linear bias model with 
$b_2=0$
is disfavored at 
68\% C.L.  
That is, the nonlinear scale-dependent bias is needed
to match the measured power spectrum, as 
can be found from Fig.~\ref{fig:PkvsBF}. 
Similar to Fig.2, bimodal structure of the constraint on $b_2$ is found, 
implying that the term proportional to $b_{2}^{2}$ in
Eq.~(\ref{eq:nonlinearPkhalo}) is dominant over other terms in the
nonlinear power spectrum. \par 

\section{Summary} 

In this paper, we explore the robustness of the PT-based model to interpret the measured galaxy 
power spectrum, focusing on constraining the neutrino mass. The model successfully include 
the effects of nonlinear clustering and nonlinear, scale-dependent galaxy bias in a self-consistent 
manner within the PT framework. We have tested the accuracy of the PT model by comparing 
the model predictions with the halo power spectrum measured in the {\it N}-body simulation without 
massive neutrinos. A careful and detailed comparison shows that the PT model can reproduce the 
simulated halo power spectrum and recover the cosmological parameters input in the simulations 
within statistical uncertainties, if the power spectrum is used up to $k\simeq 0.15\,h$Mpc$^{-1}$. 
However, in the case of the redshift-space power spectrum, the best-fit cosmological parameters 
show a biased estimation from the input values if the information up to $k\simeq 0.2\,h$Mpc$^{-1}$ is used. 
Thus, it is unclear to choose the maximum range 
of wavenumber but we decided to conservatively use the observed power spectrum up to 
$k=0.1\,h$Mpc$^{-1}$ in order to minimize possible unknown nonlinear systematic effects. 

Based on the test against the mock power spectra, we have applied this PT model to the SDSS LRG samples, 
and derived the neutrino mass limit $\sum m_\nu\le 0.81~{\rm eV}$ (95\% C.L.).  
The parameter constraints including neutrino masses would be
further improved by including the redshift distortion measurement and/or
the higher-order clustering information, which 
help to break the degeneracies with galaxy bias parameters. 
On a theory side, the PT-based 
model needs to be further refined by including higher-order loop corrections
and/or by calibrating the model with a suit of high-resolution
simulations (see \cite{Taruya:2010fj} for such a study), although 
a careful treatment of massive neutrinos, 
nonlinear galaxy bias and redshift
distortion is definitely needed. 
Once such refined models are available, 
a more stringent constraint on neutrino masses 
can be obtained from 
high-precision measurements of galaxy clustering 
via on-going and future galaxy redshift surveys. 
We hope that this paper gives the first attempt to step in this direction.

\acknowledgments
We thank I.~Kayo for allowing us to use
 his {\it N}-body simulation results, and we thank 
B.~Reid, H.~Murayama, M.~White,
and O.~Lahav for useful discussion and valuable comments. 
S.S. is supported by JSPS through the Excellent Young
Researchers Overseas Visit Program.  S.S., M.T. and A.T. are supported
in part by a Grants-in-Aid for Scientific Research from the JSPS:
for S.S., No. 21-00784; for M.T., Nos. 17740129 and 18072001; 
and for A.T., No. 21740168. 
This work is also supported in part by World
Premier International Research Center Initiative (WPI Initiative), MEXT,
Japan.

\bibliographystyle{apsrev}
\bibliography{LRGnu}

\begin{thebibliography}{29}
\expandafter\ifx\csname natexlab\endcsname\relax\def\natexlab#1{#1}\fi
\expandafter\ifx\csname bibnamefont\endcsname\relax
  \def\bibnamefont#1{#1}\fi
\expandafter\ifx\csname bibfnamefont\endcsname\relax
  \def\bibfnamefont#1{#1}\fi
\expandafter\ifx\csname citenamefont\endcsname\relax
  \def\citenamefont#1{#1}\fi
\expandafter\ifx\csname url\endcsname\relax
  \def\url#1{\texttt{#1}}\fi
\expandafter\ifx\csname urlprefix\endcsname\relax\def\urlprefix{URL }\fi
\providecommand{\bibinfo}[2]{#2}
\providecommand{\eprint}[2][]{\url{#2}}

\bibitem[{\citenamefont{{Hu} et~al.}(1998)\citenamefont{{Hu}, {Eisenstein}, and
  {Tegmark}}}]{Huetal:1998}
\bibinfo{author}{\bibfnamefont{W.}~\bibnamefont{{Hu}}},
  \bibinfo{author}{\bibfnamefont{D.~J.} \bibnamefont{{Eisenstein}}},
  \bibnamefont{and}
  \bibinfo{author}{\bibfnamefont{M.}~\bibnamefont{{Tegmark}}},
  \bibinfo{journal}{Physical Review Letters} \textbf{\bibinfo{volume}{80}},
  \bibinfo{pages}{5255} (\bibinfo{year}{1998}),
  \eprint{arXiv:astro-ph/9712057}.

\bibitem[{\citenamefont{{Takada} et~al.}(2006)\citenamefont{{Takada},
  {Komatsu}, and {Futamase}}}]{Takadaetal:2006}
\bibinfo{author}{\bibfnamefont{M.}~\bibnamefont{{Takada}}},
  \bibinfo{author}{\bibfnamefont{E.}~\bibnamefont{{Komatsu}}},
  \bibnamefont{and}
  \bibinfo{author}{\bibfnamefont{T.}~\bibnamefont{{Futamase}}},
  \bibinfo{journal}{\prd} \textbf{\bibinfo{volume}{73}},
  \bibinfo{pages}{083520} (\bibinfo{year}{2006}),
  \eprint{arXiv:astro-ph/0512374}.

\bibitem[{\citenamefont{{Elgar{\o}y} et~al.}(2002)\citenamefont{{Elgar{\o}y},
  {Lahav}, {Percival}, {Peacock}, {Madgwick}, {Bridle}, {Baugh}, {Baldry},
  {Bland-Hawthorn}, {Bridges} et~al.}}]{Elgaroy:2002lr}
\bibinfo{author}{\bibfnamefont{{\O}.}~\bibnamefont{{Elgar{\o}y}}},
  \bibinfo{author}{\bibfnamefont{O.}~\bibnamefont{{Lahav}}},
  \bibinfo{author}{\bibfnamefont{W.~J.} \bibnamefont{{Percival}}},
  \bibinfo{author}{\bibfnamefont{J.~A.} \bibnamefont{{Peacock}}},
  \bibinfo{author}{\bibfnamefont{D.~S.} \bibnamefont{{Madgwick}}},
  \bibinfo{author}{\bibfnamefont{S.~L.} \bibnamefont{{Bridle}}},
  \bibinfo{author}{\bibfnamefont{C.~M.} \bibnamefont{{Baugh}}},
  \bibinfo{author}{\bibfnamefont{I.~K.} \bibnamefont{{Baldry}}},
  \bibinfo{author}{\bibfnamefont{J.}~\bibnamefont{{Bland-Hawthorn}}},
  \bibinfo{author}{\bibfnamefont{T.}~\bibnamefont{{Bridges}}},
  \bibnamefont{et~al.}, \bibinfo{journal}{Physical Review Letters}
  \textbf{\bibinfo{volume}{89}}, \bibinfo{pages}{061301}
  (\bibinfo{year}{2002}), \eprint{arXiv:astro-ph/0204152}.

\bibitem[{\citenamefont{{Tegmark} et~al.}(2006)\citenamefont{{Tegmark},
  {Eisenstein}, {Strauss}, {Weinberg}, {Blanton}, {Frieman}, {Fukugita},
  {Gunn}, {Hamilton}, {Knapp} et~al.}}]{Tegmark:2006fk}
\bibinfo{author}{\bibfnamefont{M.}~\bibnamefont{{Tegmark}}},
  \bibinfo{author}{\bibfnamefont{D.~J.} \bibnamefont{{Eisenstein}}},
  \bibinfo{author}{\bibfnamefont{M.~A.} \bibnamefont{{Strauss}}},
  \bibinfo{author}{\bibfnamefont{D.~H.} \bibnamefont{{Weinberg}}},
  \bibinfo{author}{\bibfnamefont{M.~R.} \bibnamefont{{Blanton}}},
  \bibinfo{author}{\bibfnamefont{J.~A.} \bibnamefont{{Frieman}}},
  \bibinfo{author}{\bibfnamefont{M.}~\bibnamefont{{Fukugita}}},
  \bibinfo{author}{\bibfnamefont{J.~E.} \bibnamefont{{Gunn}}},
  \bibinfo{author}{\bibfnamefont{A.~J.~S.} \bibnamefont{{Hamilton}}},
  \bibinfo{author}{\bibfnamefont{G.~R.} \bibnamefont{{Knapp}}},
  \bibnamefont{et~al.}, \bibinfo{journal}{\prd} \textbf{\bibinfo{volume}{74}},
  \bibinfo{pages}{123507} (\bibinfo{year}{2006}),
  \eprint{arXiv:astro-ph/0608632}.

\bibitem[{\citenamefont{{Thomas} et~al.}(2010)\citenamefont{{Thomas},
  {Abdalla}, and {Lahav}}}]{Thomas:2010lr}
\bibinfo{author}{\bibfnamefont{S.~A.} \bibnamefont{{Thomas}}},
  \bibinfo{author}{\bibfnamefont{F.~B.} \bibnamefont{{Abdalla}}},
  \bibnamefont{and} \bibinfo{author}{\bibfnamefont{O.}~\bibnamefont{{Lahav}}},
  \bibinfo{journal}{Physical Review Letters} \textbf{\bibinfo{volume}{105}},
  \bibinfo{pages}{031301} (\bibinfo{year}{2010}), \eprint{0911.5291}.

\bibitem[{\citenamefont{{Reid} et~al.}(2010)\citenamefont{{Reid}, {Percival},
  {Eisenstein}, {Verde}, {Spergel}, {Skibba}, {Bahcall}, {Budavari}, {Frieman},
  {Fukugita} et~al.}}]{Reid:2010qy}
\bibinfo{author}{\bibfnamefont{B.~A.} \bibnamefont{{Reid}}},
  \bibinfo{author}{\bibfnamefont{W.~J.} \bibnamefont{{Percival}}},
  \bibinfo{author}{\bibfnamefont{D.~J.} \bibnamefont{{Eisenstein}}},
  \bibinfo{author}{\bibfnamefont{L.}~\bibnamefont{{Verde}}},
  \bibinfo{author}{\bibfnamefont{D.~N.} \bibnamefont{{Spergel}}},
  \bibinfo{author}{\bibfnamefont{R.~A.} \bibnamefont{{Skibba}}},
  \bibinfo{author}{\bibfnamefont{N.~A.} \bibnamefont{{Bahcall}}},
  \bibinfo{author}{\bibfnamefont{T.}~\bibnamefont{{Budavari}}},
  \bibinfo{author}{\bibfnamefont{J.~A.} \bibnamefont{{Frieman}}},
  \bibinfo{author}{\bibfnamefont{M.}~\bibnamefont{{Fukugita}}},
  \bibnamefont{et~al.}, \bibinfo{journal}{\mnras}
  \textbf{\bibinfo{volume}{404}}, \bibinfo{pages}{60} (\bibinfo{year}{2010}),
  \eprint{0907.1659}.

\bibitem[{\citenamefont{{Brandbyge} and {Hannestad}}(2009)}]{Brandbyge:2009}
\bibinfo{author}{\bibfnamefont{J.}~\bibnamefont{{Brandbyge}}} \bibnamefont{and}
  \bibinfo{author}{\bibfnamefont{S.}~\bibnamefont{{Hannestad}}},
  \bibinfo{journal}{Journal of Cosmology and Astro-Particle Physics}
  \textbf{\bibinfo{volume}{5}}, \bibinfo{pages}{2} (\bibinfo{year}{2009}),
  \eprint{0812.3149}.

\bibitem[{\citenamefont{{Viel} et~al.}(2010)\citenamefont{{Viel}, {Haehnelt},
  and {Springel}}}]{Vieletal:2010}
\bibinfo{author}{\bibfnamefont{M.}~\bibnamefont{{Viel}}},
  \bibinfo{author}{\bibfnamefont{M.~G.} \bibnamefont{{Haehnelt}}},
  \bibnamefont{and}
  \bibinfo{author}{\bibfnamefont{V.}~\bibnamefont{{Springel}}},
  \bibinfo{journal}{ArXiv e-prints}  (\bibinfo{year}{2010}),
  \eprint{1003.2422}.

\bibitem[{\citenamefont{{Agarwal} and {Feldman}}(2010)}]{Feldman:2010}
\bibinfo{author}{\bibfnamefont{S.}~\bibnamefont{{Agarwal}}} \bibnamefont{and}
  \bibinfo{author}{\bibfnamefont{H.~A.} \bibnamefont{{Feldman}}},
  \bibinfo{journal}{ArXiv e-prints}  (\bibinfo{year}{2010}),
  \eprint{1006.0689}.

\bibitem[{\citenamefont{{Saito} et~al.}(2008)\citenamefont{{Saito}, {Takada},
  and {Taruya}}}]{Saito:2008lr}
\bibinfo{author}{\bibfnamefont{S.}~\bibnamefont{{Saito}}},
  \bibinfo{author}{\bibfnamefont{M.}~\bibnamefont{{Takada}}}, \bibnamefont{and}
  \bibinfo{author}{\bibfnamefont{A.}~\bibnamefont{{Taruya}}},
  \bibinfo{journal}{Physical Review Letters} \textbf{\bibinfo{volume}{100}},
  \bibinfo{pages}{191301} (\bibinfo{year}{2008}), \eprint{0801.0607}.

\bibitem[{\citenamefont{{Saito} et~al.}(2009)\citenamefont{{Saito}, {Takada},
  and {Taruya}}}]{Saito:2009fk}
\bibinfo{author}{\bibfnamefont{S.}~\bibnamefont{{Saito}}},
  \bibinfo{author}{\bibfnamefont{M.}~\bibnamefont{{Takada}}}, \bibnamefont{and}
  \bibinfo{author}{\bibfnamefont{A.}~\bibnamefont{{Taruya}}},
  \bibinfo{journal}{\prd} \textbf{\bibinfo{volume}{80}},
  \bibinfo{pages}{083528} (\bibinfo{year}{2009}), \eprint{0907.2922}.

\bibitem[{\citenamefont{{Wong}}(2008)}]{Wong:2008qy}
\bibinfo{author}{\bibfnamefont{Y.~Y.~Y.} \bibnamefont{{Wong}}},
  \bibinfo{journal}{\jcap} \textbf{\bibinfo{volume}{10}}, \bibinfo{pages}{35}
  (\bibinfo{year}{2008}), \eprint{0809.0693}.

\bibitem[{\citenamefont{{Shoji} and {Komatsu}}(2009)}]{Shoji:2009fk}
\bibinfo{author}{\bibfnamefont{M.}~\bibnamefont{{Shoji}}} \bibnamefont{and}
  \bibinfo{author}{\bibfnamefont{E.}~\bibnamefont{{Komatsu}}},
  \bibinfo{journal}{\apj} \textbf{\bibinfo{volume}{700}}, \bibinfo{pages}{705}
  (\bibinfo{year}{2009}), \eprint{0903.2669}.

\bibitem[{\citenamefont{{Lesgourgues} et~al.}(2009)\citenamefont{{Lesgourgues},
  {Matarrese}, {Pietroni}, and {Riotto}}}]{Lesgourgues:2009uq}
\bibinfo{author}{\bibfnamefont{J.}~\bibnamefont{{Lesgourgues}}},
  \bibinfo{author}{\bibfnamefont{S.}~\bibnamefont{{Matarrese}}},
  \bibinfo{author}{\bibfnamefont{M.}~\bibnamefont{{Pietroni}}},
  \bibnamefont{and} \bibinfo{author}{\bibfnamefont{A.}~\bibnamefont{{Riotto}}},
  \bibinfo{journal}{\jcap} \textbf{\bibinfo{volume}{6}}, \bibinfo{pages}{17}
  (\bibinfo{year}{2009}), \eprint{0901.4550}.

\bibitem[{\citenamefont{{Jeong} and {Komatsu}}(2009)}]{JeongKomatsu:2009}
\bibinfo{author}{\bibfnamefont{D.}~\bibnamefont{{Jeong}}} \bibnamefont{and}
  \bibinfo{author}{\bibfnamefont{E.}~\bibnamefont{{Komatsu}}},
  \bibinfo{journal}{\apj} \textbf{\bibinfo{volume}{691}}, \bibinfo{pages}{569}
  (\bibinfo{year}{2009}), \eprint{0805.2632}.

\bibitem[{\citenamefont{{Nishimichi} et~al.}(2009)\citenamefont{{Nishimichi},
  {Shirata}, {Taruya}, {Yahata}, {Saito}, {Suto}, {Takahashi}, {Yoshida},
  {Matsubara}, {Sugiyama} et~al.}}]{Nishimichietal:2009}
\bibinfo{author}{\bibfnamefont{T.}~\bibnamefont{{Nishimichi}}},
  \bibinfo{author}{\bibfnamefont{A.}~\bibnamefont{{Shirata}}},
  \bibinfo{author}{\bibfnamefont{A.}~\bibnamefont{{Taruya}}},
  \bibinfo{author}{\bibfnamefont{K.}~\bibnamefont{{Yahata}}},
  \bibinfo{author}{\bibfnamefont{S.}~\bibnamefont{{Saito}}},
  \bibinfo{author}{\bibfnamefont{Y.}~\bibnamefont{{Suto}}},
  \bibinfo{author}{\bibfnamefont{R.}~\bibnamefont{{Takahashi}}},
  \bibinfo{author}{\bibfnamefont{N.}~\bibnamefont{{Yoshida}}},
  \bibinfo{author}{\bibfnamefont{T.}~\bibnamefont{{Matsubara}}},
  \bibinfo{author}{\bibfnamefont{N.}~\bibnamefont{{Sugiyama}}},
  \bibnamefont{et~al.}, \bibinfo{journal}{Publ.~Astron.~Soc.~Japan}
  \textbf{\bibinfo{volume}{61}}, \bibinfo{pages}{321} (\bibinfo{year}{2009}),
  \eprint{0810.0813}.

\bibitem[{\citenamefont{{Taruya} et~al.}(2009)\citenamefont{{Taruya},
  {Nishimichi}, {Saito}, and {Hiramatsu}}}]{Taruya:2009uq}
\bibinfo{author}{\bibfnamefont{A.}~\bibnamefont{{Taruya}}},
  \bibinfo{author}{\bibfnamefont{T.}~\bibnamefont{{Nishimichi}}},
  \bibinfo{author}{\bibfnamefont{S.}~\bibnamefont{{Saito}}}, \bibnamefont{and}
  \bibinfo{author}{\bibfnamefont{T.}~\bibnamefont{{Hiramatsu}}},
  \bibinfo{journal}{\prd} \textbf{\bibinfo{volume}{80}},
  \bibinfo{pages}{123503} (\bibinfo{year}{2009}), \eprint{0906.0507}.

\bibitem[{\citenamefont{{Taruya} et~al.}(2010)\citenamefont{{Taruya},
  {Nishimichi}, and {Saito}}}]{Taruya:2010fj}
\bibinfo{author}{\bibfnamefont{A.}~\bibnamefont{{Taruya}}},
  \bibinfo{author}{\bibfnamefont{T.}~\bibnamefont{{Nishimichi}}},
  \bibnamefont{and} \bibinfo{author}{\bibfnamefont{S.}~\bibnamefont{{Saito}}},
  \bibinfo{journal}{ArXiv e-prints}  (\bibinfo{year}{2010}),
  \eprint{1006.0699}.

\bibitem[{\citenamefont{{Komatsu} et~al.}(2009)\citenamefont{{Komatsu},
  {Dunkley}, {Nolta}, {Bennett}, {Gold}, {Hinshaw}, {Jarosik}, {Larson},
  {Limon}, {Page} et~al.}}]{Komatsu:2009qy}
\bibinfo{author}{\bibfnamefont{E.}~\bibnamefont{{Komatsu}}},
  \bibinfo{author}{\bibfnamefont{J.}~\bibnamefont{{Dunkley}}},
  \bibinfo{author}{\bibfnamefont{M.~R.} \bibnamefont{{Nolta}}},
  \bibinfo{author}{\bibfnamefont{C.~L.} \bibnamefont{{Bennett}}},
  \bibinfo{author}{\bibfnamefont{B.}~\bibnamefont{{Gold}}},
  \bibinfo{author}{\bibfnamefont{G.}~\bibnamefont{{Hinshaw}}},
  \bibinfo{author}{\bibfnamefont{N.}~\bibnamefont{{Jarosik}}},
  \bibinfo{author}{\bibfnamefont{D.}~\bibnamefont{{Larson}}},
  \bibinfo{author}{\bibfnamefont{M.}~\bibnamefont{{Limon}}},
  \bibinfo{author}{\bibfnamefont{L.}~\bibnamefont{{Page}}},
  \bibnamefont{et~al.}, \bibinfo{journal}{\apjs}
  \textbf{\bibinfo{volume}{180}}, \bibinfo{pages}{330} (\bibinfo{year}{2009}),
  \eprint{0803.0547}.

\bibitem[{\citenamefont{{Swanson} et~al.}(2010)\citenamefont{{Swanson},
  {Percival}, and {Lahav}}}]{Swanson:2010lr}
\bibinfo{author}{\bibfnamefont{M.~E.~C.} \bibnamefont{{Swanson}}},
  \bibinfo{author}{\bibfnamefont{W.~J.} \bibnamefont{{Percival}}},
  \bibnamefont{and} \bibinfo{author}{\bibfnamefont{O.}~\bibnamefont{{Lahav}}},
  \bibinfo{journal}{\mnras} \textbf{\bibinfo{volume}{409}},
  \bibinfo{pages}{1100} (\bibinfo{year}{2010}), \eprint{1006.2825}.

\bibitem[{\citenamefont{{McDonald}}(2006)}]{McDonald:2006kx}
\bibinfo{author}{\bibfnamefont{P.}~\bibnamefont{{McDonald}}},
  \bibinfo{journal}{\prd} \textbf{\bibinfo{volume}{74}},
  \bibinfo{pages}{103512} (\bibinfo{year}{2006}),
  \eprint{arXiv:astro-ph/0609413}.

\bibitem[{\citenamefont{{Tang} et~al.}(2011)\citenamefont{{Tang}, {Kayo}, and
  {Takada}}}]{Tangetal}
\bibinfo{author}{\bibfnamefont{J.}~\bibnamefont{{Tang}}},
  \bibinfo{author}{\bibfnamefont{I.}~\bibnamefont{{Kayo}}}, \bibnamefont{and}
  \bibinfo{author}{\bibfnamefont{M.}~\bibnamefont{{Takada}}}
  (\bibinfo{year}{2011}), \eprint{in preparation}.

\bibitem[{\citenamefont{{Kaiser}}(1987)}]{Kaiser:87}
\bibinfo{author}{\bibfnamefont{N.}~\bibnamefont{{Kaiser}}},
  \bibinfo{journal}{\mnras} \textbf{\bibinfo{volume}{227}}, \bibinfo{pages}{1}
  (\bibinfo{year}{1987}).

\bibitem[{\citenamefont{{Hamana} et~al.}(2003)\citenamefont{{Hamana}, {Kayo},
  {Yoshida}, {Suto}, and {Jing}}}]{Hamanaetal:03}
\bibinfo{author}{\bibfnamefont{T.}~\bibnamefont{{Hamana}}},
  \bibinfo{author}{\bibfnamefont{I.}~\bibnamefont{{Kayo}}},
  \bibinfo{author}{\bibfnamefont{N.}~\bibnamefont{{Yoshida}}},
  \bibinfo{author}{\bibfnamefont{Y.}~\bibnamefont{{Suto}}}, \bibnamefont{and}
  \bibinfo{author}{\bibfnamefont{Y.~P.} \bibnamefont{{Jing}}},
  \bibinfo{journal}{\mnras} \textbf{\bibinfo{volume}{343}},
  \bibinfo{pages}{1312} (\bibinfo{year}{2003}),
  \eprint{arXiv:astro-ph/0305187}.

\bibitem[{\citenamefont{{Percival} et~al.}(2004)\citenamefont{{Percival},
  {Verde}, and {Peacock}}}]{Percival:2004np}
\bibinfo{author}{\bibfnamefont{W.~J.} \bibnamefont{{Percival}}},
  \bibinfo{author}{\bibfnamefont{L.}~\bibnamefont{{Verde}}}, \bibnamefont{and}
  \bibinfo{author}{\bibfnamefont{J.~A.} \bibnamefont{{Peacock}}},
  \bibinfo{journal}{\mnras} \textbf{\bibinfo{volume}{347}},
  \bibinfo{pages}{645} (\bibinfo{year}{2004}), \eprint{arXiv:astro-ph/0306511}.

\bibitem[{\citenamefont{{Komatsu} et~al.}(2010)\citenamefont{{Komatsu},
  {Smith}, {Dunkley}, {Bennett}, {Gold}, {Hinshaw}, {Jarosik}, {Larson},
  {Nolta}, {Page} et~al.}}]{Komatsu:2010yq}
\bibinfo{author}{\bibfnamefont{E.}~\bibnamefont{{Komatsu}}},
  \bibinfo{author}{\bibfnamefont{K.~M.} \bibnamefont{{Smith}}},
  \bibinfo{author}{\bibfnamefont{J.}~\bibnamefont{{Dunkley}}},
  \bibinfo{author}{\bibfnamefont{C.~L.} \bibnamefont{{Bennett}}},
  \bibinfo{author}{\bibfnamefont{B.}~\bibnamefont{{Gold}}},
  \bibinfo{author}{\bibfnamefont{G.}~\bibnamefont{{Hinshaw}}},
  \bibinfo{author}{\bibfnamefont{N.}~\bibnamefont{{Jarosik}}},
  \bibinfo{author}{\bibfnamefont{D.}~\bibnamefont{{Larson}}},
  \bibinfo{author}{\bibfnamefont{M.~R.} \bibnamefont{{Nolta}}},
  \bibinfo{author}{\bibfnamefont{L.}~\bibnamefont{{Page}}},
  \bibnamefont{et~al.}, \bibinfo{journal}{ArXiv e-prints}
  (\bibinfo{year}{2010}), \eprint{1001.4538}.

\bibitem[{\citenamefont{{Eisenstein} et~al.}(2005)\citenamefont{{Eisenstein},
  {Zehavi}, {Hogg}, {Scoccimarro}, {Blanton}, {Nichol}, {Scranton}, {Seo},
  {Tegmark}, {Zheng} et~al.}}]{Eisensteinetal:2005}
\bibinfo{author}{\bibfnamefont{D.~J.} \bibnamefont{{Eisenstein}}},
  \bibinfo{author}{\bibfnamefont{I.}~\bibnamefont{{Zehavi}}},
  \bibinfo{author}{\bibfnamefont{D.~W.} \bibnamefont{{Hogg}}},
  \bibinfo{author}{\bibfnamefont{R.}~\bibnamefont{{Scoccimarro}}},
  \bibinfo{author}{\bibfnamefont{M.~R.} \bibnamefont{{Blanton}}},
  \bibinfo{author}{\bibfnamefont{R.~C.} \bibnamefont{{Nichol}}},
  \bibinfo{author}{\bibfnamefont{R.}~\bibnamefont{{Scranton}}},
  \bibinfo{author}{\bibfnamefont{H.}~\bibnamefont{{Seo}}},
  \bibinfo{author}{\bibfnamefont{M.}~\bibnamefont{{Tegmark}}},
  \bibinfo{author}{\bibfnamefont{Z.}~\bibnamefont{{Zheng}}},
  \bibnamefont{et~al.}, \bibinfo{journal}{\apj} \textbf{\bibinfo{volume}{633}},
  \bibinfo{pages}{560} (\bibinfo{year}{2005}), \eprint{arXiv:astro-ph/0501171}.

\bibitem[{\citenamefont{{Percival} et~al.}(2010)\citenamefont{{Percival},
  {Reid}, {Eisenstein}, {Bahcall}, {Budavari}, {Frieman}, {Fukugita}, {Gunn},
  {Ivezi{\'c}}, {Knapp} et~al.}}]{Percival:2010lr}
\bibinfo{author}{\bibfnamefont{W.~J.} \bibnamefont{{Percival}}},
  \bibinfo{author}{\bibfnamefont{B.~A.} \bibnamefont{{Reid}}},
  \bibinfo{author}{\bibfnamefont{D.~J.} \bibnamefont{{Eisenstein}}},
  \bibinfo{author}{\bibfnamefont{N.~A.} \bibnamefont{{Bahcall}}},
  \bibinfo{author}{\bibfnamefont{T.}~\bibnamefont{{Budavari}}},
  \bibinfo{author}{\bibfnamefont{J.~A.} \bibnamefont{{Frieman}}},
  \bibinfo{author}{\bibfnamefont{M.}~\bibnamefont{{Fukugita}}},
  \bibinfo{author}{\bibfnamefont{J.~E.} \bibnamefont{{Gunn}}},
  \bibinfo{author}{\bibfnamefont{{\v Z}.}~\bibnamefont{{Ivezi{\'c}}}},
  \bibinfo{author}{\bibfnamefont{G.~R.} \bibnamefont{{Knapp}}},
  \bibnamefont{et~al.}, \bibinfo{journal}{\mnras}
  \textbf{\bibinfo{volume}{401}}, \bibinfo{pages}{2148} (\bibinfo{year}{2010}),
  \eprint{0907.1660}.

\bibitem[{\citenamefont{{Lewis} and {Bridle}}(2002)}]{Lewis:2002lr}
\bibinfo{author}{\bibfnamefont{A.}~\bibnamefont{{Lewis}}} \bibnamefont{and}
  \bibinfo{author}{\bibfnamefont{S.}~\bibnamefont{{Bridle}}},
  \bibinfo{journal}{\prd} \textbf{\bibinfo{volume}{66}},
  \bibinfo{pages}{103511} (\bibinfo{year}{2002}),
  \eprint{arXiv:astro-ph/0205436}.

\end{thebibliography}

\end{document}